\documentclass[11pt]{article}
\usepackage{amssymb}

\usepackage{graphicx}
\usepackage{amsmath}
\usepackage{makeidx}
\usepackage{indentfirst}

\newcounter{resultnum}[section]

\newcounter{conclusionnum}[section]

\newcounter{conditionnum}[section]

\newcounter{conjecturenum}[section]

\newcounter{examplenum}[section]

\newcounter{exercisenum}[section]

\newcounter{lemmanum}[section]

\newcounter{notationnum}[section]

\newcounter{theoremnum}[section]

\newcounter{definitionnum}[section]

\newcounter{corollarynum}[section]

\newcounter{remarknum}[section]

\newcounter{propositionnum}[section]

\newcounter{acknowledgementnum}[section]

\newcounter{algorithmnum}[section]

\newcounter{axiomnum}[section]

\newcounter{casenum}[section]

\newcounter{claimnum}[section]

\newcounter{summarynum}[section]

\newcounter{problemnum}[section]

\begin{document}

\title{Nonholonomic Black Ring and Solitonic Solutions in Finsler and Extra Dimension Gravity Theories}
\date{April 30, 2010}
\author{\textbf{Mihai Anastasiei}\thanks{%
anastas@uaic.ro} \\
%EndAName
\textsl{{\small Faculty of Mathematics, University "Al. I. Cuza" Ia\c{s}i,}}
\\
%EndAName
\textsl{{\small 11, Carol I Boulevard, Ia\c{s}i, Romania, 700506 }} \\
%EndAName
\textsl{\small and } \\
%EndAName
\textsl{{\small Mathematical Institute "O. Mayer", Romanian Academy Ia\c{s}i
Branch,}} \\
%EndAName
\textsl{{\small 8, Carol I Boulevard, Ia\c{s}i, Romania, 700506 } } \and
%EndAName
\textbf{Sergiu I. Vacaru} \thanks{%
sergiu.vacaru@uaic.ro, Sergiu.Vacaru@gmail.com \newline
http://www.scribd.com/people/view/1455460-sergiu} \and \textsl{\small %
Science Department, University "Al. I. Cuza" Ia\c si, } \\
%EndAName
\textsl{\small 54, Lascar Catargi street, Ia\c si, Romania, 700107 } }
\maketitle

\begin{abstract}
We study stationary configurations mimicking nonholonomic locally an\-isotropic black rings (for instance, with ellipsoidal polarizations and/or imbedded into solitonic backgrounds) in three/six dimensional pseudo--Finsler/ Riemannian spacetimes. In the asymptotically flat limit,
for holonomic configurations, a subclass of such spacetimes contains the set of five dimensional black ring solutions with regular rotating event horizon. For corresponding parameterizations, the metrics and connections define Finsler--Einstein geometries modeled on tangent bundles, or on nonholonomic (pseudo) Riemannian manifolds. In general, there are vacuum nonholonomic gravitational configurations which can not be generated in the limit of zero cosmological constant.

\vskip0.1cm



\end{abstract}

%\tableofcontents

%\newpage

\section{Introduction}

There is a recent interest in (pseudo) Finsler geometry and applications to
gravity \cite{perl,mignemi,gibbons,sindoni,skak,vncg,vsgg}, cosmology and
astrophysics \cite{vgont,lin,chang,kstav}, see reviews of results and
methods in Refs. \cite{ijgmmp,vrflg}. This paper is a partner of work \cite%
{vfbh}, where Finsler black hole, ellipsoid and solitonic solutions were
constructed for two classes of models of (pseudo) Finsler gravity on
nonholonomic (pseudo) Riemannian manifolds and/or tangent bundles. We follow
our anholonomic deformation method of constructing exact solutions in gravity and
the aim of this letter is to study axisymmetric stationary solutions in
higher dimensions and show how black ring configurations can be generated in
Finsler gravity theories.\footnote{%
The solutions analyzed in this work are different from the locally
anisotropic (black) ellipsod/torus configurations considered in Chapters
10-12 of \ Ref. \cite{vsgg}. In this paper, our goal is to analyze black
ring metrics and their nonholonomic deformations just for the (pseudo)
Finsler spaces.}

The most important property of the axisymmetric stationary solutions both in
Finsler like theories and extra--dimension gravity\ is the fact that they
admit event horizons with non--spherical topology which is in contrast to
the four dimensional general relativity theory. In general, the topology of
the event horizon can not be uniquely determined which provides a number of
possible theoretical and experimental verifications of gravity theories and
analogous models of classical and quantum interactions. In order to consider
nonholonomic transforms of exact solutions in (pseudo) Finsler and/or extra
dimension gravity with nontrivial topology and possible spacetime topology
transitions, we shall use as 'prime'' metrics certain classes of six
dimensional metrics containing as imbedding five--dimensional black
ring/hole configurations, for instance, having topology of $\mathbb{S}^{3}$%
--sphere, or $\mathbb{S}^{1}\times \mathbb{S}^{2}$--torus.\footnote{%
in this paper, we shall use the term spacetime for a (pseudo)
Riemannian/Finsler manifold enabled with a metric structure of signature $%
\pm $} The 'target' metrics generated by nonholonomic deforms will be
constructed to possess the same, or different type topology, which for
Finsler spaces positively can be more complicated and with a more 'rich'
spacetime geometry because of existence of a nontrivial nonlinear connection
(N--connection) structure.\footnote{%
We consider that our readers are familiar with the main geometric concepts
of Lagrange--Finsler geometry modelled on tangent bundles \cite{ma1987,ma},
or on (pseudo) Riemannian manifolds enabled with nonintegrable distributions
(i.e. nonholonomic/N--anholonomic manifolds) \cite{vfbh,vrflg,vsgg}.}

Assuming the existence of two additional commutating axial killing vector
fields and the horizon topology of black ring $\mathbb{S}^{1}\times \mathbb{S%
}^{2},$ it was found \cite{mty} that there is only one asymptotically flat
black ring solution with a regular horizon which is the so--called
Pomeransky--Sen'kov black ring \cite{ps}. There were also found many other
black ring/object solutions (black Saturn/ torus--ellipsoid configurations,
di--ring/ bi--ring etc), see reviews and original results in Refs. \cite%
{er,tz,ar,ch,rog1} and Part II of monograph \cite{vsgg} where solutions with
torus and ellipsoid nonholonomic configurations are investigated in details.

The discoveries that certain uniqueness black hole and non--hair theorems
are violated in higher dimensions were regarded as very surprising and
related to solutions with non--spherical horizon topology. We add to the
list of such counterexamples of black hole fundamental theorems a new series
of solutions generated by nonholonomic deformations and modeling Finsler
configurations on higher dimension Einstein spaces and/or tangent bundles.

The paper is organized as follows: In section 2, we fix a very general
ansatz for a class of metrics with toroidal topology which are
nonholonomically deformed on (pseudo) Finsler/ Riemannian spacetimes. Such
metrics are subjected to the condition to define Finsler--Einstein
spacetimes as exact solutions of corresponding Einstein equations with
cosmological constant. Section 3 is devoted to a class of 6-dimensional
exact solutions with nontrivial cosmological constants. There are considered
certain conditions when such solutions define nonholonomic deformations of
black ring metrics and (pseudo) Finsler polarized black rings with possible
ellipsoidal deformations and solitonic perturbations. There are stated the
conditions when Finsler type solutions transform into the Levi--Civita ones.
In section 4, we analyze (pseudo) Finsler stationary vacuum solutions for
the so--called canonical distinguished connection in Finsler geometry and
the Levi--Civita connection in extra dimension gravity. We show that for
small nonholonomic deformations, it is possible to generate Finsler type
black ring solutions with small polarizations of physical parameters.
Finally, we provide some concluding remarks in section 5.

\section{Nonhlonomic Ring Ansatz for Finsler--Einstein Spaces}

\subsection{Geometric preliminaries}

We can consider a six dimensional (6-d) manifold \ $\mathbf{V}$ of necessary
differentiability class. We endow it with a $3$--dimensional non--integrable
distribution $\mathcal{D}.$ The pair $(\mathbf{V,}\mathcal{D})$ is called a
nonholonomic manifold. We label the local coordinates $u=(x,y)$ on an open
region $U\subset \mathbf{V}$ in the form $u^{\alpha }=(x^{i},y^{a})$ with
indices $i,j,k,...=1,2,3$ and $a,b,c...=4,5,6.$\footnote{%
In a similar form we shall use 'primed' indices and coordinates, $u^{\alpha
^{\prime }}=(x^{i^{\prime }},y^{a^{\prime }}),$ when $i^{\prime },j^{\prime
},k^{\prime },...=1,2,3$ and $a^{\prime },b^{\prime },c^{\prime }...=4,5,6.$}
Let $e_{i^{\prime}}$ be a local frame for $\mathcal{D}$ $\ $such that $e_{i^{\prime
}}=e_{\ i^{\prime }}^{i}\frac{\partial }{\partial x^{i}}+e_{\ i^{\prime
}}^{a}\frac{\partial }{\partial y^{a}}$ with $\det |e_{\ i^{\prime
}}^{i}|\neq 0.$ The Einstein convention on summation is applied. For any set
of frame coefficients $e_{\ i^{\prime }}^{i},$ we can define $e_{i\ }^{\
j^{\prime }}\equiv (e^{-1})_{\ j^{\prime }}^{i}$ $\ $following formulas $%
e_{\ i^{\prime }}^{i}e_{i\ }^{\ j^{\prime }}=\delta _{i^{\prime
}}^{j^{\prime }},$ where $\delta _{i^{\prime }}^{j^{\prime }}$ is the
Kronecker symbol (such frame and coframe coefficients can be defined, in
general, on any $U).$

Expressing $\frac{\partial }{\partial x^{i}}=e_{i\ }^{\ j^{\prime
}}e_{j^{\prime }}+e_{i}^{\ a^{\prime }}\frac{\partial }{\partial
y^{a^{\prime }}},$ taking $\mathbf{e}_{i}=$ $e_{i\ }^{\ j^{\prime
}}e_{j^{\prime }}$ as a new local frame on $\mathcal{D}$ considering that $%
e_{a}^{\ a^{\prime }}=$ $\delta _{a}^{a^{\prime }}$ and $e_{i}^{\ a}\equiv
N_{i}^{a}(x,y),$ we get $\mathbf{e}_{i}=\delta _{i}=\frac{\partial }{%
\partial x^{i}}-N_{i}^{a}(u)\frac{\partial }{\partial y^{a}}.$\footnote{%
In our works, boldface symbols are used for nonholonomic manifolds/ bundles
(nonholonomic spaces) and geometric objects on such spaces; we can consider
that $\delta _{i}$ are the so--called N--elongated partial derivatives in
Finsler geometry and generalizations.} We may complete $\mathbf{e}_{i}$ to a
local frame $\mathbf{e}_{\alpha }=(\mathbf{e}_{i},e_{a}),$ for $%
e_{a}=\partial _{a}\equiv \frac{\partial }{\partial y^{a}},$ on $\mathbf{V.}$
This fact suggests to consider also on $\mathbf{V}$ the distribution $%
\widetilde{\mathcal{D}}$ locally spanned by $\partial _{a}.$ It is
supplementary to $\mathcal{D}$ and locally integrable. If a change of
coordinates $(x^{i},y^{a})$ $\rightarrow (x^{i^{\prime
}}(x^{i},y^{a}),y^{a^{\prime }}(x^{i},y^{a}))$ is performed, the formula  $%
\frac{\partial }{\partial y^{a}}=\frac{\partial x^{i^{\prime }}}{\partial
y^{a}}\frac{\partial }{\partial x^{i^{\prime }}}+\frac{\partial y^{a^{\prime
}}}{\partial y^{a}}\frac{\partial }{\partial y^{a^{\prime }}}$ has to
simplify to $\frac{\partial }{\partial y^{a}}=\frac{\partial y^{a^{\prime }}%
}{\partial y^{a}}\frac{\partial }{\partial y^{a^{\prime }}}$ and thus $\frac{%
\partial x^{i^{\prime }}}{\partial y^{a}}=0.$ This equality and the general
condition $\delta _{j}=e_{j\ }^{\ j^{\prime }}\delta _{j^{\prime }}$ with $%
\det |e_{j\ }^{\ j^{\prime }}|\neq 0,$ where \ $\delta _{j^{\prime }}=\frac{%
\partial }{\partial x^{j^{\prime }}}-N_{j^{\prime }}^{a^{\prime }}(u)\frac{%
\partial }{\partial y^{a^{\prime }}},$ imply $e_{j\ }^{\ j^{\prime }}=\frac{%
\partial x^{j^{\prime }}}{\partial x^{j}}$ and $\frac{\partial x^{j^{\prime
}}}{\partial x^{j}}N_{j^{\prime }}^{a^{\prime }}=\frac{\partial y^{a^{\prime
}}}{\partial y^{a}}N_{j}^{a}-\frac{\partial y^{a^{\prime }}}{\partial x^{j}}.
$ Alternatively, we may say that $\mathbf{V}$ is a foliated manifold with
foliation determined by the distribution $\widetilde{\mathcal{D}},$ called also
structural distribution, and having $\mathcal{D}$ as a transversal
distribution.

In this work the nonholonomic aspects will be more important and the stress
will be upon the distribution $\mathcal{D}.$ In particular, $\mathbf{V}$
can be the total spaces, for instance, of a submersion over a 3--d manifold $%
M,$ or (more particular) a vector bundle $E$ over $M,$ a principal bundle
over $M,$ or the (co) tangent bundle $TM$ ($T^{\ast }M).$ The general
geometrical framework just described, terms and techniques from Finsler and Lagrange
geometries \cite{ma1987,ma,vsgg,ijgmmp,vrflg,bm} will be used. Thus we call
a decomposition $T\mathbf{V=}\mathcal{D\oplus }\widetilde{\mathcal{D}}$ as a
nonlinear connection (N--connection) structure $\mathbf{N}$ on $\mathbf{V}$
with local coefficients $N_{i}^{a}(u).$\footnote{%
in this work, we shall use the so--called canonical Cartan N--connection and
the distinguished connection, d--connection, see details in Refs. \cite%
{ma,vrflg}} For $\mathbf{V=}TM\mathbf{,}$ we can write $TTM\mathbf{=}%
hTM\oplus hTM$ with certain natural ''horizontal'' (h) and ''vertical'' (v)
decompositions. In such a case, a N--connection structure can be derived
canonically from any regular Lagrangian $L:TM\rightarrow \mathbb{R}$ (or for
some more particular homogeneous on variables $y^{a}$ cases, from a
fundamental (generating) Finsler function $F(x,y),$ when $F(x,\beta y)=\beta
F(x,y),\beta >0).$ This allows us to construct various models of
Lagrange--Finsler geometry and gravity theories on $TM,$ when the variables $%
y^{a}$ are of ''velocity/momentum'' type. Alternatively, we can work on a
(pseudo) Riemannian nonholonomic manifold  $\mathbf{V}$ endowed with a
corresponding metric structure $\mathbf{g}=\{g_{\alpha \beta }(u)\}.$ For
this class of  nonholonomic geometry and gravity theories, any types of
conventional distributions  $\mathcal{D}$ and $\widetilde{\mathcal{D}}$ can
be introduced on $\mathbf{V}$ on convenience, for instance, with the aim to
define a general geometric method of constructing exact solutions in general
relativity and various modifications of  Ricci flow theory \cite%
{vbep,vs5dbh,vrf1,vsgg,ijgmmp,vrflg}, when $y^{a}$ can be treated as certain
general nonholonomically constrained variables/coordinates (they can be
defined even on usual Einstein spaces of arbitrary dimension).

\subsection{Ring type Einstein and Finsler--Einstein spaces}

We shall use a 'prime' metric on $\mathbf{V}$
\begin{eqnarray}
\mathbf{\check{g}} &=&\check{g}_{i}(x^{k})dx^{i}\otimes dx^{i}+\check{h}%
_{a}(x^{k})\mathbf{\check{e}}^{a}\otimes \mathbf{\check{e}}^{a},
\label{targdm} \\
\mathbf{\check{e}}^{a} &=&dy^{a}+\check{N}_{i}^{a}(x^{k})dx^{i}.  \notag
\end{eqnarray}%
For certain parameterizations \footnote{$\check{g}_{1}=-1,\check{g}_{2}=%
\check{g}_{2}(x^{2},x^{3}),\check{g}_{3}=\check{g}_{3}(x^{2},x^{3}),$ $%
\check{h}_{6}(x^{k})=\pm 1$ and some vanishing N--connection coefficients
with respect to a corresponding local coordinate basis, $\ \check{N}%
_{i}^{4}=0,\check{N}_{i}^{6}=0,$ $\check{N}_{2,3}^{5}=0,$ but $\check{N}%
_{i}^{5}\neq 0,$ and for local coordinates $x^{i}=x^{i}(\psi
,x,y),y^{4}=\phi ,y^{5}=t,y^{6}=y^{6}$}, this metric contains as a trivial
embedding (with extension on coordinate $y^{6},$ for a term of type $\pm
(dy^{6})^{2}$ in the metric quadratic form, into a 6--d pseudo--Riemannian
space) the 5--d black ring metric with a dipole, or with rotation in the $%
\mathbb{S}^{2}$ \cite{figueras}. The 5-d part of (\ref{targdm}), in
coordinates $(x^{i},y^{4},y^{5}),$ is prescribed to be conformally
proportional, with the factor $\ (x-y)^{2}F(y)/\check{R}^{2}F(x)G(y),$ to
the black ring metric
\begin{eqnarray}
^{br}\mathbf{g} &=&\ \check{R}^{2}\frac{F(x)}{(x-y)^{2}}\left[ -\frac{G(y)}{%
F(y)}d\psi \otimes d\psi +G^{-1}(x)dx\otimes dx-G^{-1}(y)dy\otimes dy\right]
\notag \\
&&+\check{R}^{2}\frac{G(x)}{(x-y)^{2}}d\phi \otimes d\phi  \label{targdm1} \\
&&-\frac{F(y)}{F(x)}\left( dt-C\check{R}\frac{1+y}{F(y)}d\psi \right)
\otimes \left( dt-C\check{R}\frac{1+y}{F(y)}d\psi \right) .  \notag
\end{eqnarray}%
The coefficients of this metric are stated by a class of functions and
constants: for instance, $F(x)=1+\check{\lambda}x$ and $G(y)=(1-y^{2})(1+\nu
y)$ for $\check{R}=const$ and $C=\sqrt{\check{\lambda}(\check{\lambda}-\nu
)(1+\check{\lambda})/(1-\check{\lambda})}$ is determined by certain
dimensionless parameters $\check{\lambda}$ and $\nu $ satisfying the
condition $0<\nu \leq \check{\lambda}<1$ (here we note that our notations
are different from those used in Ref. \cite{er}). The metric (\ref{targdm1})
defines an exact solution of Einstein equations for the the Levi--Civita in
5-d gravity.

Using nonholonomic deformations of the locally isotropic black ring metric (%
\ref{targdm}) with $g_{i}(x^{\hat{k}})=\eta _{i}(x^{\hat{k}})\check{g}_{i}$
and $h_{\widehat{a}}=\eta _{\widehat{a}}(x^{k},y^{4})\check{h}_{\widehat{a}\
}$ and some nontrivial $N_{i}^{4}=w_{i}(x^{k},y^{4})$ and $%
N_{i}^{5}=n_{i}(x^{k},y^{4}),$ but $N_{i}^{6}=0,$ when indices with ''hats''
run values $\widehat{i},\hat{k},...=2,3$ and $\widehat{a},\widehat{b}%
,...=4,5,$ we generate a metric of type\
\begin{eqnarray}
\mathbf{g} &=&g_{i}(x^{\hat{k}})e^{i}\otimes e^{j}+h_{a}(x^{k},y^{4})\mathbf{%
e}^{a}\otimes \mathbf{e}^{b},  \label{gpsm} \\
\mathbf{e}^{a} &=&dy^{a}+N_{i}^{a}(x^{k},y^{4})dx^{i},  \label{dfram}
\end{eqnarray}%
when values $\mathbf{g}_{\alpha ^{\prime }\beta ^{\prime }}=[g_{i},h_{a}]$ \
with $g_{1}=-1$ and $h_{6}=\pm 1$ are related by frame transforms
\begin{equation}
\mathbf{g}_{\alpha ^{\prime }\beta ^{\prime }}e_{\ \alpha }^{\alpha ^{\prime
}}(x^{k},y^{a})e_{\ \beta }^{\beta ^{\prime }}(x^{k},y^{a})=\mathbf{f}%
_{\alpha \beta }(x^{k},y^{a})  \label{algeq}
\end{equation}%
to a (pseudo--Finsler) metric $\ \mathbf{f}_{\alpha \beta }=\ [\ f_{ij},\
f_{ab}]$ and corresponding N--adapted dual canonical basis $\ ^{c}\mathbf{e}%
^{\alpha }=\left( dx^{i},\ ^{c}\mathbf{e}^{a}=dy^{a}+\
^{c}N_{i}^{a}(x^{k},y^{a})dx^{i}\right) $ defined by the so--called
canonical Cartan N--connection in Finsler geometry, see details in Refs. %
\cite{vfbh,vrflg,ijgmmp,ma1987,ma}.\footnote{\label{fnfg} Any (pseudo)
Finsler metric$\ \mathbf{f=\{f}$ $_{\alpha \beta }\}$ can be parametrized in
the canonical Sasaki form
\begin{equation*}
\ \mathbf{f}=\ f_{ij}dx^{i}\otimes dx^{j}+\ f_{ab}\ ^{c}\mathbf{e}%
^{a}\otimes \ ^{c}\mathbf{e}^{b},\ \ ^{c}\mathbf{e}^{a}=dy^{a}+\
^{c}N_{i}^{a}dx^{i},
\end{equation*}%
where the (pseudo) Finsler configuration is defied by 1) a fundamental real
Finsler (generating) function $F(u)=F(x,y)=F(x^{i},y^{a})>0$ if $y\neq 0$
and homogeneous of type $F(x,\lambda y)=|\lambda |F(x,y),$ for any nonzero $%
\lambda \in \mathbb{R},$ with positively definite Hessian $\ f_{ab}=\frac{1}{%
2}\frac{\partial ^{2}F^{2}}{\partial y^{a}\partial y^{b}},$ when $\det |\
f_{ab}|\neq 0.$ The \ Cartan canonical N--connection structure $\ ^{c}%
\mathbf{N}=\{\ ^{c}N_{i}^{a}\}$ is completely defined by an effective
Lagrangian $L=F^{2}$ in such a form that the corresponding semi--spray
configuration is defined by nonlinear geodesic equations being equivalent to
the Euler--Lagrange equations for $L$ (see details, for instance, in Refs. %
\cite{ma,vrflg,vsgg}; for ''pseudo'' configurations, this mechanical analogy
is a formal one, with some ''imaginary'' coordinates \cite{vfbh}). One
defines $\ \ ^{c}N_{i}^{a}=\frac{\partial G^{a}}{\partial y^{3+i}}$ \ for $%
G^{a}=\frac{1}{4}\ f^{a\ 3+i}\left( \frac{\partial ^{2}L}{\partial
y^{3+i}\partial x^{k}}y^{3+k}-\frac{\partial L}{\partial x^{i}}\right) ,$
where $\ f^{ab}$ is inverse to $\ f_{ab}$ and respective contractions of
horizontal (h) and vertical (v) indices, $\ i,j,...$ and $a,b...,$ are
performed following the rule: we can write, for instance, an up $v$--index $%
a $ as $a=3+i$ and contract it with a low index $i=1,2,3.$ In brief, for
spaces of even dimension, we shall write $y^{i}$ instead of $y^{3+i},$ or $%
y^{a}.$} For any given values $\mathbf{g}_{\alpha ^{\prime }\beta ^{\prime
}} $ and $~\mathbf{f}_{\alpha \beta },$ we have to solve a system of
quadratic algebraic equations (\ref{algeq}) in order to determine the
unknown variables $e_{\ \alpha }^{\alpha ^{\prime }}.$ How to define in
explicit form such frame \ coefficients (vierbeins) and coordinates in 4--d
we discuss in Refs. \cite{vfbh,ijgmmp}, but the algebraic computations are
similar for 6--d. As a matter of principle, any (pseudo) Riemannian metric $%
\mathbf{g}_{\alpha ^{\prime }\beta ^{\prime }}$ can be expressed in Finsler
variables as a metric $\mathbf{f}_{\alpha \beta },$ up to a trivial
imbedding into an even dimension, (the inverse statement also holds true),
if a N--connection structure is prescribed on $\mathbf{V.}$ The standard
(pseudo) Finsler geometric and gravitational models (for instance, those
from \cite{ma1987,ma}) are constructed on $\mathbf{V}=TM,$ when 'fiber'
variables $y^{a}$ are treated as velocities, but Finsler like structures/
variables can be defined also in the Einstein gravity and string
generalizations \cite{vrflg,ijgmmp,vsgg} when $y^{a}$ are considered as
certain nonholonomically constrained coordinates on (pseudo) Riemannian
manifolds with possible torsion generalizations.

In this work, we shall analyze a class of 6--d metrics (or 3--d Finsler
metrics) defining Finsler--Einstein spaces as exact solutions of the
Einstein equations,
\begin{equation}
\widehat{R}_{\ j}^{i}=\ \lambda \delta _{\ j}^{i},\ \widehat{R}_{\
b}^{a}=\lambda \delta _{\ b}^{a},\ \widehat{R}_{ia}=\ \widehat{R}_{ai}=0
\label{eeqcdcc}
\end{equation}%
where $\widehat{\mathbf{R}}_{\alpha \beta }=\{\widehat{R}_{ij},\widehat{R}%
_{ia},\widehat{R}_{ai},\widehat{R}_{ab}\}$ are the components of the Ricci
tensor computed for the canonical distinguished connection (d--connection) $%
\ \widehat{\mathbf{D}},$ see details in \cite{vfbh,vrflg,ijgmmp,vsgg}%
\footnote{%
The coefficients of $\widehat{\mathbf{D}}$ with respect to a
''N--elongated'' frames of type (\ref{dfram}) are computed $\widehat{\mathbf{%
\Gamma }}_{\ \alpha \beta }^{\gamma }=\left( \widehat{L}_{jk}^{i},\widehat{L}%
_{bk}^{a},\widehat{C}_{jc}^{i},\widehat{C}_{bc}^{a}\right) $ with
\begin{eqnarray*}
\widehat{L}_{jk}^{i} &=&\frac{1}{2}g^{ir}\left(
e_{k}g_{jr}+e_{j}g_{kr}-e_{r}g_{jk}\right) , \\
\widehat{L}_{bk}^{a} &=&e_{b}(N_{k}^{a})+\frac{1}{2}h^{ac}\left(
e_{k}h_{bc}-h_{dc}\ e_{b}N_{k}^{d}-h_{db}\ e_{c}N_{k}^{d}\right) , \\
\widehat{C}_{jc}^{i} &=&\frac{1}{2}g^{ik}e_{c}g_{jk},\ \widehat{C}_{bc}^{a}=%
\frac{1}{2}h^{ad}\left( e_{c}h_{bd}+e_{c}h_{cd}-e_{d}h_{bc}\right) .
\end{eqnarray*}%
Any geometric construction for the canonical d--connection $\widehat{\mathbf{%
D}}$ can be re--defined equivalently into a similar one with the
Levi--Civita connection $\bigtriangledown =\{\ _{\shortmid }\Gamma _{\beta
\gamma }^{\alpha }\},$ and inversely following the formula $\ _{\shortmid
}\Gamma _{\ \alpha \beta }^{\gamma }=\widehat{\mathbf{\Gamma }}_{\ \alpha
\beta }^{\gamma }+\ _{\shortmid }Z_{\ \alpha \beta }^{\gamma },$ where the
distortion tensor $\ _{\shortmid }Z_{\ \alpha \beta }^{\gamma }$ is computed
as
\begin{eqnarray*}
\ _{\shortmid }Z_{jk}^{a} &=&-C_{jb}^{i}g_{ik}h^{ab}-\frac{1}{2}\Omega
_{jk}^{a},~_{\shortmid }Z_{bk}^{i}=\frac{1}{2}\Omega
_{jk}^{c}h_{cb}g^{ji}-\Xi _{jk}^{ih}~C_{hb}^{j}, \\
_{\shortmid }Z_{bk}^{a} &=&~^{+}\Xi _{cd}^{ab}~~^{\circ }L_{bk}^{c},\
_{\shortmid }Z_{kb}^{i}=\frac{1}{2}\Omega _{jk}^{a}h_{cb}g^{ji}+\Xi
_{jk}^{ih}~C_{hb}^{j},\ _{\shortmid }Z_{jk}^{i}=0, \\
\ _{\shortmid }Z_{jb}^{a} &=&-~^{-}\Xi _{cb}^{ad}~~^{\circ }L_{dj}^{c},\
_{\shortmid }Z_{bc}^{a}=0,\ _{\shortmid }Z_{ab}^{i}=-\frac{g^{ij}}{2}\left[
~^{\circ }L_{aj}^{c}h_{cb}+~^{\circ }L_{bj}^{c}h_{ca}\right] ,
\end{eqnarray*}%
for $\ \Xi _{jk}^{ih}=\frac{1}{2}(\delta _{j}^{i}\delta
_{k}^{h}-g_{jk}g^{ih}),~^{\pm }\Xi _{cd}^{ab}=\frac{1}{2}(\delta
_{c}^{a}\delta _{d}^{b}+h_{cd}h^{ab})$ and$~^{\circ
}L_{aj}^{c}=L_{aj}^{c}-e_{a}(N_{j}^{c}).$}, $\ \lambda $ is a cosmological
constant, and $\delta _{\ j}^{i},$ for instance, is the Kronecker symbol.
Solutions of \ nonholonomic equations (\ref{eeqcdcc}) \ are typical ones for
the Finsler gravity with metric compatible d--connections (in our case, $%
\widehat{\mathbf{D}}\mathbf{f=0)}$ modelled in various types of (super)
string, noncommutative and other generalizations, see reviews of results in
references in \cite{vncg,vsgg,vrflg}.

Imposing additional nonholonomic constraints on coefficients of metrics
solving the system (\ref{eeqcdcc}) but when the distorsion of the Ricci
tensor (under nonholonomic deforms from one linear connection to another) is
zero, we select a more restricted class of Finsler--Einstein configurations
defining exact solutions in the 6--d generalization of the Einstein gravity,
with the Ricci tensor $\ _{\shortmid }R_{\alpha \beta }$ for the
Levi--Civita connection $\nabla ,$ i.e. solutions of
\begin{equation}
\ _{\shortmid }R_{\alpha \beta }=\lambda g_{\alpha \beta }.  \label{eeqlcc}
\end{equation}
There are Finsler--Einstein configurations with $\widehat{\mathbf{R}}%
_{\alpha \beta }=\ _{\shortmid }R_{\alpha \beta }$ but, in general, such
spaces have different curvature tensors because they are defined by
different linear connections.

The goal of this work is to construct and analyze possible physical
implications of a class of exact solutions of equations (\ref{eeqcdcc}) and (%
\ref{eeqlcc}) with $\lambda \neq 0,$ or $\lambda =0,$ with metrics of type (%
\ref{gpsm}) containing in certain limits the black ring metric (\ref{targdm1}%
). The \ new classes of nonholonomically deformed black \ rings will be
considered for the (pseudo) Finsler gravity on tangent bundle and/or for
usual 6--d Einstein spacetimes.

\section{6--d Solutions with Nontrivial Cosmological Constant}

In this section we consider two classes of exact solutions of (non)
holonomic Einstein equations with $\lambda \neq 0$ generated by
corresponding conformal transform of the 5--d black ring metric (\ref%
{targdm1}), nonholonomic deformations and imbedding into a 6-d (pseudo)
Riemannian, or in a 3--d (pseudo) Finsler spacetime.

\subsection{Ansatz for prime and target metrics}

Let us consider a prime metric%
\begin{eqnarray}
_{cf}^{br}\mathbf{g} &=&-d\psi \otimes d\psi +\frac{(x-y)^{2}F(y)}{\check{R}%
^{2}F(x)G(y)}\left[ \frac{dx\otimes dx}{G(x)}-\frac{dy\otimes dy}{G(y)}%
\right]  \notag \\
&&+\frac{^{1}G(x)F(y)}{F(x)G(y)}d\phi \otimes d\phi  \label{ctbrm} \\
&&-\frac{(x-y)^{2}F^{2}(y)}{\check{R}^{2}F^{2}(x)G(y)}\left( dt-C\check{R}%
\frac{1+y}{F(y)}d\psi \right) \otimes \left( dt-C\check{R}\frac{1+y}{F(y)}%
d\psi \right) ,  \notag
\end{eqnarray}%
which multiplied to the conformal factor $\ \check{R}^{2}\frac{F(x)}{%
(x-y)^{2}}\frac{G(y)}{F(y)}$ and for $\ ^{1}G(x)=G(x)$ is just the black
ring metric (\ref{targdm1}). We introduce variables
\begin{equation*}
x^{1}=\psi ,x^{2}=\int dx/\sqrt{|G(x)|},x^{3}=\int dy/\sqrt{|G(y)|}%
,y^{4}=\phi ,y^{5}=t,y^{6}=y^{6}
\end{equation*}%
and label the metric coefficients as%
\begin{eqnarray*}
\check{g}_{2}(x^{2},x^{3}) &=&\frac{\left[ x(x^{2})-y(x^{3})\right]
^{2}F(x^{3})}{\check{R}^{2}F(x^{2})G(x^{3})}=-\check{g}_{3}(x^{2},x^{3}),%
\check{g}_{1}=-1,\ \check{h}_{6}=\pm 1, \\
\check{h}_{4}(x^{2},x^{3}) &=&\frac{\ ^{1}G(x^{2})\ F(x^{3})}{%
F(x^{2})G(x^{3})},\ \check{h}_{5}(x^{2},x^{3})=-\frac{\left[
x(x^{2})-y(x^{3})\right] ^{2}\ F(x^{3})}{\check{R}^{2}F(x^{2})G(x^{3})}, \\
\check{N}_{1}^{5}(x^{3}) &=&C\check{R}\frac{1+y(x^{3})}{F(x^{3})},\check{N}%
_{2}^{5}=\check{N}_{3}^{5}=0,\check{N}_{i}^{4}=0,\check{N}_{i}^{6}=0.
\end{eqnarray*}%
Imbedding $_{cf}^{br}\mathbf{g}$ into a 6--d spacetime, we get a metric
\begin{eqnarray}
~\mathbf{\check{g}} &=&\check{g}_{1}dx^{1}\otimes dx^{1}+\check{g}_{2}\
dx^{2}\otimes dx^{2}+\check{g}_{3}\ dx^{3}\otimes dx^{3}  \notag \\
&&+\check{h}_{4}\mathbf{\check{e}}^{4}\otimes \mathbf{\check{e}}^{4}+\check{h%
}_{5}\mathbf{\check{e}}^{5}\otimes \mathbf{\check{e}}^{5}+\check{h}_{6}%
\mathbf{\check{e}}^{6}\otimes \mathbf{\check{e}}^{6},  \notag \\
\mathbf{\check{e}}^{4} &=&dy^{4},\mathbf{\check{e}}^{5}=dy^{5}+\check{N}%
_{1}^{5}(x^{3})dx^{1},\mathbf{\check{e}}^{6}=dy^{6},  \label{trgm}
\end{eqnarray}%
which, in general, is not a solution of field equations for a gravitational
model. We search for some classes of metrics generated by nonholonomic
deformations with ''polarization'' multiples $\eta _{\alpha }=[\eta
_{i}(u^{\beta }),\eta _{a}(u^{\beta })],$ when $g_{\alpha }=\eta _{\alpha }%
\check{g}_{\alpha }=[g_{i}=\eta _{i}\check{g}_{i},h_{a}=\eta _{a}\check{h}%
_{a}],$ and modified N--connection coefficients $N_{i}^{a}(u^{\beta })$
resulting in exact solutions of (\ref{eeqcdcc}) and/or (\ref{eeqlcc}).

For
\begin{eqnarray*}
\eta _{1} &=&1,\eta _{2}\check{g}_{2}=\epsilon _{2}e^{\underline{\phi }%
(x^{2},x^{3})},\eta _{3}\check{g}_{3}=\epsilon _{3}e^{\underline{\phi }%
(x^{2},x^{3})}, \\
\eta _{6} &=&1,h_{4}\left( x^{k},y^{4}\right) =\eta _{4}\check{h}%
_{4},h_{5}\left( x^{k},y^{4}\right) =\eta _{5}\check{h}_{5},
\end{eqnarray*}%
where $\epsilon _{\alpha }=\pm 1$ for chosen signatures, and
\begin{equation*}
N_{i}^{4}=w_{i}\left( x^{k},{y}^{4}\right) ,N_{i}^{5}=n_{i}\left( x^{k},{y}%
^{4}\right) ,N_{i}^{6}=0,
\end{equation*}%
we get a class of 'target' generic off--diagonal metrics\footnote{%
which can not be diagonalized, in general, by any coordinate transform}
\begin{eqnarray}
~~^{\lambda }\mathbf{g} &=&-dx^{1}\otimes dx^{1}+e^{\underline{\varphi }%
(x^{2},x^{3})}[\epsilon _{2}\ dx^{2}\otimes dx^{2}+\epsilon _{3}\
dx^{3}\otimes dx^{3}]  \label{lambsol} \\
&&+h_{4}\left( x^{k},y^{4}\right) \ ~\mathbf{e}^{4}\otimes ~\mathbf{e}%
^{4}+h_{5}\left( x^{k},y^{4}\right) \ ~\mathbf{e}^{5}\otimes ~\mathbf{e}%
^{5}\pm dy^{6}\otimes dy^{6},  \notag \\
\mathbf{e}^{4} &=&d{y}^{4}+w_{i}\left( x^{k},{y}^{4}\right) dx^{i},\ ~%
\mathbf{e}^{5}=d{y}^{5}+n_{i}\left( x^{k},{y}^{4}\right) dx^{i},  \notag
\end{eqnarray}%
with the coefficients will be defined in next sections such a way that these
metrics are exact solutions of the Einstein equations in (pseudo)
Riemannian/ Finsler gravity.

\subsection{(Pseudo) Finsler polarized black rings}

We shall use brief denotations for some partial derivatives like $a^{\bullet
}=\partial a/\partial x^{2},$\ $a^{\prime }=\partial a/\partial x^{3},$\ $%
a^{\ast }=\partial a/\partial y^{4}.$

\subsubsection{Stationary nonholonmic deformations of black ring metrics}

By straightforward computations\footnote{%
for simplicity, we omit such computations in this work which are similar to
those presented in section 2.7, pages 143-150, in Ref. \cite{vsgg}, see also
reviews on the anholonomic defortaion method of constructing exact \ solutions in %
\cite{vncg,ijgmmp,vrflg}}, we can verify that a metric of type (\ref{lambsol}%
) generates an exact solution of the Einstein equations (\ref{eeqcdcc}) if
the coefficients are determined by any functions satisfying (respectively)
the conditions:%
\begin{eqnarray}
&&\epsilon _{2}\underline{\varphi }^{\bullet \bullet }(x^{\widehat{k}%
})+\epsilon _{3}\underline{\varphi }^{^{\prime \prime }}(x^{\widehat{k}%
})=-2\epsilon _{2}\epsilon _{3}\ \lambda ;  \label{coeflsoldc} \\
h_{4} &=&\pm \frac{\left( \varphi ^{\ast }\right) ^{2}}{4\ \lambda }e^{-2\
^{0}\varphi (x^{i})},\ h_{5}=\mp \frac{1}{4\ \lambda }e^{2(\varphi -\
^{0}\varphi (x^{i}))};  \notag \\
w_{i} &=&-\partial _{i}\varphi /\varphi ^{\ast };  \notag \\
n_{i} &=&\ ^{1}n_{i}(x^{k})+\ ^{2}n_{i}(x^{k})\int dy^{4}\left( \varphi
^{\ast }\right) ^{2}e^{-2(\varphi -\ ^{0}\varphi (x^{i}))}=  \notag \\
&&\ \left\{
\begin{array}{rcl}
^{1}n_{i}(x^{k})+\ ^{2}n_{i}(x^{k})\int dy^{4}e^{-4\varphi }\frac{\left(
h_{5}^{\ast }\right) ^{2}}{h_{5}}, & \mbox{ if \ } & n_{i}^{\ast }\neq 0; \\
\ ^{1}n_{i}(x^{k}),\quad \qquad \qquad \qquad \qquad \qquad & \mbox{ if \ }
& n_{i}^{\ast }=0;%
\end{array}%
\right.  \notag
\end{eqnarray}%
for any nonzero $h_{\widehat{a}}$ and $h_{\widehat{a}}^{\ast }$ and
(integration) functions $^{1}n_{i}(x^{k}),\ ^{2}n_{i}(x^{k}),$ a generating
function $\varphi (x^{i},y^{4}),$ and $\ ^{0}\varphi (x^{i})$ to be
determined from certain boundary conditions for a fixed system of
coordinates with $x^{\widehat{k}}=\{x^{2},x^{3}\}.$

There are two classes of solutions (\ref{coeflsoldc}) constructed for a
nontrivial $\lambda .$ The first one is singular for $\lambda \rightarrow 0$
if we choose a generation function $\varphi (x^{i},y^{4})$ not depending on $%
\lambda .$ It is possible to eliminate such singularities for certain
parametric dependencies of type $\varphi (\lambda ,x^{i},y^{4}),$ when the
resulting metric and N--connection coefficients are not singular on $\lambda
.$

\subsubsection{Small eccentricity ellipsoid polarizatons and ring
deformations}

It is not clear what kind of physical implications one may have exact
solutions with general coefficients of type (\ref{coeflsoldc}). But it is
possible to extract a subclass of solutions decomposed on a small parameter $%
\varepsilon$ which will define certain small nonholonomic deformations of
the conformally transformed black ring metric (\ref{ctbrm}) on a 6--d
(pseudo) Riemannian space.\footnote{%
Such a solution will be an exact solution because this type of series
decompositions are not on coordinate variables but for certain fixed small
parameters which can be always defined for metrics with Killing symmetries,
see details in Ref. \cite{vncg,ijgmmp,vrflg}.}

We chose a polarization function (i.e. a nonlinear self--polarization of
gravitational vacuum with cosmological constant $\lambda $) of type
\begin{equation}
\eta _{4}=\eta (x^{i},\phi ,\varepsilon ),  \label{pol1}
\end{equation}%
where $\eta (x^{i},\phi ,\varepsilon )$ is any linear on $\varepsilon $
function depending in general on variables $x^{i}$ and $\phi .$ Introducing $%
h_{4}=\eta _{4}\check{h}_{4}$ and $h_{5}=\eta _{5}\check{h}_{5}$ into the
second line of equations (\ref{coeflsoldc}), integrating on $\phi $ and
stating the integration function $\ ^{0}\varphi (x^{i})=1$, we get
\begin{eqnarray}
\varphi &=&2\sqrt{|\lambda \check{h}_{4}|}\int d\phi \sqrt{|\eta (x^{i},\phi
,\varepsilon )|},  \label{pol2} \\
\eta _{5} &=&\mp \frac{1}{4\lambda \check{h}_{5}}\exp \left[ 4\sqrt{|\lambda
\check{h}_{4}|}\int d\phi \sqrt{|\eta (x^{i},\phi ,\varepsilon )|}\right] .
\notag
\end{eqnarray}%
Haven defined polarizations $\eta _{4}$ and $\eta _{5},$ we can compute the
coefficients $h_{4}$ and $h_{5},$ which can be used for computing the
nontrivial N--connection coefficients%
\begin{equation*}
w_{i}=-\partial _{i}\varphi /\varphi ^{\ast }\mbox{ and }n_{i}=\
^{1}n_{i}(x^{k})+\ ^{2}n_{i}(x^{k})\int \left( \varphi ^{\ast }\right)
^{2}e^{-2\varphi }d\phi .
\end{equation*}%
We choose $\ ^{1}n_{i}(x^{k})=C\check{R}\frac{1+y(x^{3})}{F(x^{3})}$ in
order to have certain similarity with the prime metric (\ref{ctbrm}).

Further approximations are possible, for instance, for $\omega
(x^{2},x^{3})=\omega _{0}x^{2}$ and $\eta (x^{i},\phi ,\varepsilon )$ when
\begin{equation*}
\eta _{4}=1+\varepsilon \widetilde{\eta }_{4}(x^{i},\phi ,\varepsilon ),\eta
_{5}=\left[ 1+\varepsilon \cos (\omega _{0}x^{2})\right] \left[
1+\varepsilon \widetilde{\eta }_{5}(x^{i},\phi ,\varepsilon )\right]
\end{equation*}%
and $\eta _{2}\check{g}_{2}=\epsilon _{2}e^{\underline{\phi }%
(x^{2},x^{3})},\eta _{3}\check{g}_{3}=\epsilon _{3}e^{\underline{\phi }%
(x^{2},x^{3})}$ are determined by any solution $\underline{\phi }$ of the
first equation in (\ref{coeflsoldc}) and $\widetilde{\eta }_{5}$ is computed
using (\ref{pol2}). This generates a subclass of Finsler--Einstein spaces
parametrized by metrics of type
\begin{eqnarray}
~~_{\varepsilon }^{\lambda }\mathbf{g} &=&-dx^{1}\otimes dx^{1}+e^{%
\underline{\varphi }(x^{2},x^{3})}[\epsilon _{2}\ dx^{2}\otimes
dx^{2}+\epsilon _{3}\ dx^{3}\otimes dx^{3}]  \notag \\
&&+\left[ 1+\varepsilon \widetilde{\eta }_{4}\right] \ \check{h}_{4}~\mathbf{%
e}^{4}\otimes ~\mathbf{e}^{4}  \notag \\
&&+\left[ 1+\varepsilon \cos (\omega _{0}x^{2})\right] \left[ 1+\varepsilon
\widetilde{\eta }_{5}\right] \ ~\check{h}_{5}\mathbf{e}^{5}\otimes ~\mathbf{e%
}^{5}\pm dy^{6}\otimes dy^{6},  \notag \\
\mathbf{e}^{4} &=&d{y}^{4}+w_{i}\left( x^{k},{y}^{4}\right) dx^{i},\ ~%
\mathbf{e}^{4}=d{y}^{5}+n_{i}\left( x^{k},{y}^{4}\right) dx^{i}.
\label{finsbr}
\end{eqnarray}%
The metric (\ref{finsbr}) with coefficients computed with respect to the
dual frame of reference $\mathbf{e}^{\alpha }=(\eta _{i}dx^{i},\mathbf{e}%
^{4},\sqrt{|\eta _{5}|}\mathbf{e}^{5},dy^{6})$ is similar to a
nonholonomically polarized 5--d black ring imbedded self--consistently in a
6--dimensional spacetime. Really, the multiple $\left[ 1+\varepsilon \cos
(\omega _{0}x^{2})\right] $ before $\check{h}_{5}$ determines an elliptic
polarization with eccentricity $\varepsilon $ when
\begin{equation}
\check{r}\sim \check{R}/[1+\varepsilon \cos (\omega _{0}x^{2})]
\label{elconf}
\end{equation}%
describes an ellipse with coordinate/ anisotropy $(\omega _{0}x^{2}).$ The
multiple $\left[ 1+\varepsilon \widetilde{\eta }_{4}\right] $ before $\check{%
h}_{4}$ can be included as a polarization of function $\ ^{1}G$ in (\ref%
{ctbrm}), when $\ ^{1}G(x^{2})\rightarrow \ G(x^{2})\eta _{4}(x^{i},\phi
,\varepsilon )\sim G\left[ 1+\varepsilon \widetilde{\eta }_{4}\right] .$
Such small on $\varepsilon $ nonholonomic deformations of the thin black
rings seem to be stable but may result in unstable solutions for minimally
spinning and/or fat black rings, see details for holonomic ring
configurations in Ref. \cite{eev}.\footnote{%
the conditions of stability for nonholonomic ring configurations should be
analyzed separately for a fixed class of nonholonomic distributions like in
the case of black ellipsoids provided, for instance, in Refs. \cite%
{vbep,vncg}}

One should emphasize that for small $\varepsilon $ there are preserved the
same type of singularities of solutions and horizons as for the usual black
ring metrics but with certain polarizations of constants and additional
smooth terms, for instance, to the curvature tensor and other
tensors/connections, all being proportional to $\varepsilon .$ More general
nonholonomic deformations result not only in possible instabilities but, in
general, in various types of ''non--ring'' stationary configurations.

\subsubsection{Solitonic perturbations of nonholonomic black rings}

It is possible to consider self--consistent \ imbedding of the metric (\ref%
{finsbr}) into nontrivial backrounds similarly as it was done for solitonic
''motions'' of black holes in five dimensional gravity \cite{vs5dbh} and/or
in nonholonomic Ricci flows of exact solutions in gravity \cite{vrf1}. There
are very different classes of solutions for nonholonomic black ring --
solitonic conigurations: 1) black rings are on ''huge'' stationary solitonic
configurations, when the resulting general solution is not of ring type and
2) certain ''small'' stationary solitonic distributions preserve the black
ring character for such generic off--diagonal metrics.

In the first case, for (\ref{pol1}), we can use a static three dimensional
solitonic distribution $\eta (x^{2},x^{3},\phi )$ defined as a solution of
solitonic equation\footnote{%
we can introduce instead such an equation any type of three/two dimensional
solitonic or other nonlinear wave equations}
\begin{equation}
\eta ^{\bullet \bullet }+\epsilon (\eta ^{\prime }+6\eta \ \eta ^{\ast
}+\eta ^{\ast \ast \ast })^{\ast }=0,\ \epsilon =\pm 1.  \label{soleq}
\end{equation}%
This induces solitonic polarizations for (\ref{pol2}) and N--connection
coefficients $w_{i}$ and $n_{i}$ obtained by integrating on $\phi $ some
types of functionals depending of $\varphi (x^{2},x^{3},\phi )$ and defining
solitonic distributions. The resulting metric is of type (\ref{finsbr}) is
constructed as a stationary superposition of 3--d solitonic distributions
for 5--d and 6--d Finsler--Einstein spacetimes when some multiples in the
metric coefficients originate from former black ring coefficients of
metrics. We need to perform a more special analysis on stability and
physical meaning of such solutions for more special cases of distributions
(this is out of scope of our paper). Here we emphasize that it is obvious
that, in general, certain solitonic hierarchies and related conservation
laws can be always associated (both for nonholonomic pseudo--Riemannian and
pseudo--Finsler configurations), see details in Refs. \cite{vsolh1,vsolh2}.

In the second case, considering that $\eta \sim 1+\varepsilon \widetilde{%
\eta }_{4}(x^{2},x^{3},\phi ,\varepsilon ),$ where $\widetilde{\eta }_{4}$
is a solution of equation (\ref{soleq}), we preserve the black ring
character of the class of solutions even some physical parameters became
solitonically polarized as we discuss below in section \ref{spbrp}. Such
solutions in (pseudo) Finsler gravity have positively physical
interpretations as black rings with locally anisotropic polarizations of the
metric coefficients, relevant constants and parameters. But in this case,
not only such polarizations distinguish these classes of solutions. The
anisotropic coordinate $\phi $ is of ''fiber' type in a tangent bundle, i.e.
a special type of velocity--coordinate. The existence of such Finsler type
black rings would be a topological evidence for broken local Lorentz
invariance. This class of solutions is not for the Levi--Civita connection,
but for the canonical d--connection. Such Finsler--Einstein spaces with
nonholonomically deformed Lorentz invariance (and broken local special
relativity theory, double special relativity models etc \cite%
{mignemi,gibbons,kstav}) can be described in terms of canonical quadratic
metric forms and N-- and d--connection structures defined for a model of
Finsler geometry on tangent bundle, as we show in footnote \ref{fnfg} and
section \ref{spfc}.

Finally, in this section, we emphasize that if we model a pseudo--Finsler
geometry on a pseudo--Riemannian 6--d spacetime with the ''fiber''
coordinates treated as extra--dimensional extensions of a 3--d gravity
model, we get an Einstein--Cartan 6--d spacetime with an effective torsion
induced by some off--diagonal metric terms and as a corresponding
nonholonomic frame effect. All geometric and physical constructions can be
redefined in terms of the Levi--Civita connection, but nevertheless, the
solutions will be those for the canonical d--connection. The above class of
solutions, for such models, will describe certain possible nontrivial
topological evidences, for instance, nonholonomic black rings, induced from
extra--dimension gravity.

\subsection{Nonholonomic deformations with the Levi--Civita connection}

It is possible to constrain the integral varieties of solutions of \ the
Einstein equations for the canonical d--connection (\ref{eeqcdcc}),
parametrized by an ansatz (\ref{lambsol}), or (\ref{finsbr}), in such a way
that for certain values of coefficients the corresponding metric will be
also a solution\ of the equations (\ref{eeqlcc}) for the Levi--Civita
connection. This is possible with respect to certain classes of N--adapted
bases when the distortion of the Ricci tensor vanishes for certain
nonholonomic configurations (see details on such constructions in Refs. \cite%
{vfbh,vrflg,ijgmmp}).

By straightforward computations, we can verify that we can extract from the
above mentioned ansatz an exact 4--d solution (contained as a trivial
imbedding in 6--d) in Einstein gravity with cosmological constant $\lambda $
if the coefficients of the metric and N--connection are subjected to the
conditions:
\begin{eqnarray}
(2e^{2\varphi }\varphi -\lambda )\left( \varphi ^{\ast }\right) ^{2}
&=&0,\varphi \neq 0,\varphi ^{\ast }\neq 0;  \label{lcls} \\
w_{2}w_{3}\left( \ln |\frac{w_{2}}{w_{3}}|\right) ^{\ast } &=&w_{3}^{\bullet
}-w_{2}^{\prime },w_{\widehat{i}}^{\ast }\neq 0,w_{1}=0;  \notag \\
w_{3}^{\bullet }-w_{2}^{\prime } &=&0,\ \mbox{ if \  }w_{i}^{\ast }=0%
\mbox{
and \  }w_{1}=0;  \notag \\
\ ^{1}n_{2}^{\prime }(x^{k})-\ ^{1}n_{3}^{\bullet }(x^{k}) &=&0,%
\mbox{ if \
}n_{i}^{\ast }=0\mbox{ and \  }n_{1}=0,  \notag
\end{eqnarray}%
which holds for any $\varphi (x^{i},y^{4})=\ln |h_{5}^{\ast }/\sqrt{%
|h_{4}h_{5}|}|=const$ \ if we include configurations with $\varphi ^{\ast
}=0.$

We can consider metrics of type (\ref{lambsol}) with coefficients of class\ (%
\ref{lcls}) formally extended to 5-d and 6-d with certain nontrivial values
of $g_{1}$ and/or $h_{5}$ which will contain as an imbedding the black ring
metric (\ref{targdm1}) and its conformal transforms on variables $x$ and $y$
and various types of nonholonomic deformations. Such solutions describe
certain nonholonomic black ring configurations for small on $\varepsilon $
deformations like in (\ref{finsbr}). Nevertheless, they are different from
those considered in \cite{mty,ps,er,tz,ar,ch,rog1,figueras,eev}. In our
case, the nonholonmic distributions are nontrivial, and induced by a nonzero
cosmological constant, which polarize physical parameters and may state, for
instance, a stationary solitonic background.

\subsection{(Pseudo) Finsler configurations}

\label{spfc} For a general (pseudo) Finsler configuration, the class of
metrics (\ref{lambsol}) with coefficients (\ref{coeflsoldc}) expressed in
Finsler variables as a metric $\mathbf{f}_{\alpha \beta },$ see formulas (%
\ref{algeq}), depends formally on all six variables $x^{i}$ and $y^{a}$ and
the imbedding into a 6--d (pseudo) Riemannian spacetime is not trivial.
Nevertheless, this class of Finsler--Einstein spaces contains two Killing
vectors because in certain systems of coordinates such metrics do not depend
on variables $y^{5}$ and $y^{6}$ in explicit form. Such metrics are
stationary because the coefficients do not depend on variable $y^{5}=t.$

For any $\mathbf{f}_{\alpha \beta }=\ \mathbf{e}_{\ \alpha }^{\alpha
^{\prime }}\ \mathbf{e}_{\ \beta }^{\beta ^{\prime }}\mathbf{g}_{\alpha
^{\prime }\beta ^{\prime }}$ corresponding to a canonical Finsler type
parametrizaton of metric and connections, see footnote \ref{fnfg}, we can
write in explicit h-- and v--components
\begin{eqnarray}
\ f_{ij} &=&e_{\ i}^{i^{\prime }}e_{\ j}^{j^{\prime }}g_{i^{\prime
}j^{\prime }}\mbox{\ and  \ }~\ f_{ab}=e_{\ a}^{a^{\prime }}e_{\
b}^{b^{\prime }}\ g_{a^{\prime }b^{\prime }},  \label{auxeq1} \\
\ N_{i^{\prime }}^{a^{\prime }} &=&e_{i^{\prime }}^{\ i}e_{\ a}^{a^{\prime
}}\ ^{c}N_{i}^{a},\mbox{\ or \ }\ ^{c}N_{i}^{a}=e_{i}^{\ i^{\prime }}e_{\
a^{\prime }}^{a}\ N_{i^{\prime }}^{a^{\prime }},  \label{auxeq2}
\end{eqnarray}%
were, for instance, $e_{a^{\prime }\ }^{\ a}$ is inverse to $e_{\
a}^{a^{\prime }}.$ We can chose $g_{i^{\prime }j^{\prime
}}=diag[g_{1^{\prime }},g_{2^{\prime }},g_{3^{\prime }}],$ $h_{a^{\prime
}b^{\prime }}=diag[h_{4^{\prime }},h_{5^{\prime }},\epsilon _{6^{\prime }}]$
and $N_{i^{\prime }}^{a^{\prime }}=\left( N_{i^{\prime }}^{3^{\prime
}}=w_{i^{\prime }},N_{i^{\prime }}^{4^{\prime }}=n_{i^{\prime
}},N_{i^{\prime }}^{5^{\prime }}=0\right) $ to be defined by any exact
solution of type (\ref{lambsol}), or (\ref{finsbr}). The (pseudo) Finsler
data $\ f_{ij},$ $\ f_{ab}$ and $\ ^{c}N_{i}^{a}$ $=\left( \ ^{c}N_{i}^{3}=\
^{c}w_{i},\ ^{c}N_{i}^{4}=\ ^{c}n_{i},0\right) $ are with diagonal matrices,
$\ f_{ij}=diag[f_{1},f_{2},f_{3}]$ and $\ f_{ab}=diag[f_{4},f_{5},\epsilon
_{6}],$ if the generating function is of type $F=\ ^{1}F(x^{i},y^{4})$ $+\
^{2}F(x^{i},y^{5})$ for some homogeneous (respectively, on $y^{4}$ and $%
y^{5})$ functions $\ ^{1}F$ and $\ ^{2}F.$\footnote{%
Of course, we can work with arbitrary generating functions $F(x^{i},y^{a})$
but this will result in off--diagonal (pseudo) Finsler metrics in N--adapted
bases, which would request a more cumbersome matrix calculus.}

The conditions (\ref{auxeq1}) are satisfied for a diagonal representation
for $\mathbf{e}_{\ \alpha }^{\alpha ^{\prime }}$ if%
\begin{eqnarray*}
e_{\ 1}^{1^{\prime }} &=&\pm \sqrt{\left| \frac{\ f_{1}}{g_{1^{\prime }}}%
\right| },e_{\ 2}^{2^{\prime }}=\pm \sqrt{\left| \frac{\ f_{2}}{g_{2^{\prime
}}}\right| }\ ,e_{\ 3}^{3^{\prime }}=\pm \sqrt{\left| \frac{\ f_{3}}{%
g_{3^{\prime }}}\right| }, \\
e_{\ 4}^{4^{\prime }} &=&\pm \sqrt{\left| \frac{\ f_{4}}{h_{4^{\prime }}}%
\right| },e_{\ 5}^{5^{\prime }}=\pm \sqrt{\left| \frac{\ f_{5}}{h_{5^{\prime
}}}\right| ,}e_{\ 6}^{6^{\prime }}=\pm 1.
\end{eqnarray*}%
For any fixed values $\ f_{i},\ f_{a}$ and $\ ^{c}w_{i},^{c}n_{i}$ and given
$g_{i^{\prime }}$ and $h_{a^{\prime }},$ we can compute $w_{i^{\prime }}$
and $\ n_{i^{\prime }}$ as
\begin{eqnarray*}
w_{1^{\prime }} &=&\pm \sqrt{\left| \frac{g_{1^{\prime }}\ f_{4}}{%
h_{4^{\prime }}\ f_{1}}\right| }\ ^{c}w_{1},w_{2^{\prime }}=\pm \sqrt{\left|
\frac{g_{2^{\prime }}\ f_{4}}{h_{4^{\prime }}\ f_{2}}\right| }\
^{c}w_{2},w_{3^{\prime }}=\pm \sqrt{\left| \frac{g_{3^{\prime }}\ f_{4}}{%
h_{4^{\prime }}\ f_{3}}\right| }\ ^{c}w_{3}, \\
n_{1^{\prime }} &=&\pm \sqrt{\left| \frac{g_{1^{\prime }}\ f_{5}}{%
h_{5^{\prime }}\ f_{1}}\right| }\ ^{c}n_{1},n_{2^{\prime }}=\pm \sqrt{\left|
\frac{g_{2^{\prime }}\ f_{5}}{h_{5^{\prime }}\ f_{2}}\right| }\
^{c}n_{2},n_{3^{\prime }}=\pm \sqrt{\left| \frac{g_{3^{\prime }}\ f_{5}}{%
h_{5^{\prime }}\ f_{3}}\right| }\ ^{c}n_{3},
\end{eqnarray*}%
which defines solutions for equations (\ref{auxeq2}).

So, any black ring metric and various types of (non) holonomic deformations
can be expressed as a Sasaki type metric for a 3--d (pseudo) Finsler
spacetime.\footnote{%
In order to generate a homogeneous model on a total tangent bundle, we have
to use another types of lifts of metrics from a base, see Ref. \cite{bm}.
For simplicity, in this work, we consider gravitational models with Sasaki
type lifts when the homogeneity is considered as a property only for the
generating function but not obligatory for other geometric objects. It is
possible to construct Finsler models when all fundamental geometric objects
are rigorously subjected to the condition of homogeneity but they are very
"nonlinear" in nature and subjected to more sophisticate conditions of
nonholonomic constraints.} If the coefficients of the nonholonomically
deformed metric of type (\ref{lambsol}), or (\ref{finsbr}), are constrained
to satisfy the conditions (\ref{coeflsoldc}), we model stationary
Finsler--Einstein spacetimes enabled with canonical d--connection structure.
As a matter of principle, we can introduce Finsler like variables on any
(pseudo) Riemannian spacetime, see details in Refs. \cite{vfbh,vrflg,ijgmmp}%
. In this case, we generate Einstein spacetimes with generic off--diagonal
metrics, equivalently described in terms both of the Levi--Civita connection
and the canonical d--connection with respect to the corresponding N--adapted
frames, if the coefficients of such exact solutions are of type (\ref{lcls}).

\section{Vacuum (Pseudo) Finsler Ring Metrics}

In this section, we construct exact solutions of equations (\ref{eeqcdcc}),
or (\ref{eeqlcc}), with $\lambda =0,$ defining vacuum (pseudo) Finsler
models of ring solutions.

\subsection{Stationary vacuum ansatz}

\subsubsection{Canonical d--connection vacuum configurations}

In general, because of generic nonholonomic and nonlinear character of
Finsler type configurations, such metrics can not be obtained in a limit $%
\lambda \rightarrow 0$ for coefficients (\ref{coeflsoldc}). It is necessary
to apply the anholonomic deformation method of generating exact solutions from the
very beginning to the vacuum Einstein equations for the canonical
d--connection, $\widehat{\mathbf{R}}_{\alpha \beta }=0,$ for an ansatz of
type $\mathbf{g}$ (\ref{lambsol}). The nontrivial coefficients of such an
exact solution must satisfy the conditions%
\begin{eqnarray}
&&\epsilon _{2}\underline{\varphi }^{\bullet \bullet }(x^{\widehat{k}%
})+\epsilon _{3}\underline{\varphi }^{^{\prime \prime }}(x^{\widehat{k}})=0;
\label{vacdsolc} \\
h_{4} &=&\pm e^{-2\ ^{0}\varphi }\frac{\left( h_{5}^{\ast }\right) ^{2}}{%
h_{5}}\mbox{ for a given }h_{5}(x^{i},y^{4}),\ \varphi =\ ^{0}\varphi =const;
\notag \\
w_{i} &=&w_{i}(x^{k},y^{4}),\mbox{ for any such functions if }\lambda =0;
\notag \\
n_{i} &=&\left\{
\begin{array}{rcl}
\ \ ^{1}n_{i}(x^{k})+\ ^{2}n_{i}(x^{k})\int \left( h_{5}^{\ast }\right)
^{2}|h_{5}|^{-5/2}dy^{4}, & \mbox{ if \ } & n_{i}^{\ast }\neq 0; \\
\ ^{1}n_{i}(x^{k}),\quad \qquad \qquad \qquad \qquad \qquad & \mbox{ if \ }
& n_{i}^{\ast }=0.%
\end{array}%
\right.  \notag
\end{eqnarray}%
Metrics of this class define vacuum nonholonomic deformations of black rings
if the coefficients of the primary ring metric $\mathbf{\check{g}}$ are
included into a target vacuum metric as $g_{i}=\eta _{i}\check{g}_{i},$ $%
h_{4}=\eta _{4}\check{h}_{4}$ and $h_{5}=\eta _{5}\check{h}_{5}$ and $\
^{1}n_{i}(x^{k})=$ $C\check{R}\frac{1+y(x^{3})}{F(x^{3})}.$

It is not clear, in general, what physical importance may have solutions
with data (\ref{vacdsolc}). In a similar manner as in the previous section,
it is possible to extract nonholonomically polarized vacuum black ring
solutions using decompositions on a small parameter $\varepsilon $ like we
have considered for the coefficients of ansatz (\ref{finsbr}). Such vacuum
black rings can be also nonholonomically imbedded into solitonic backrounds
of type (\ref{soleq}) and redefined in Finsler variables folowing transforms
of type (\ref{auxeq1}) and (\ref{auxeq2}). On tangent bundles, such
solutions are for the (pseudo) Finsler gravity with the canonical
d--connection. This class of black rings contain some nonholonomic
configurations which are topologically nontrivial and with brocken local
Lorentz symmetry. They can be also constructed in higher dimension
extensions of (pseudo) Riemannian spaces but in those cases the anisotropic
coordinates will not be of ''velocity'' type.

\subsubsection{Levi--Civita vacuum configurations}

It is necessary to impose additional constraints on the integral varieties
in order to generate exact solutions of the Einstein equations for the
Levi--Civita connection, i.e of $\ _{\shortmid }R_{\alpha \beta }=0.$ Such
additional constraints can be of type
\begin{eqnarray}
h_{4} &=&\pm 4\left[ \left( \sqrt{|h_{5}|}\right) ^{\ast }\right] ^{2},\quad
h_{5}^{\ast }\neq 0;  \notag \\
w_{2}w_{3}\left( \ln |\frac{w_{2}}{w_{3}}|\right) ^{\ast } &=&w_{3}^{\bullet
}-w_{2}^{\prime },\quad w_{i}^{\ast }\neq 0;  \notag \\
w_{3}^{\bullet }-w_{2}^{\prime } &=&0,\mbox{\  if \  }\quad w_{i}^{\ast }=0%
\mbox{ \ and \  }\quad w_{1}=0;  \notag \\
\ ^{1}n_{2}^{\prime }(x^{k})-\ ^{1}n_{3}^{\bullet }(x^{k}) &=&0,\quad %
\mbox{\  if \  }n_{1}=0,  \label{vaclcsoc}
\end{eqnarray}%
for $e^{-2\ ^{0}\varphi }=1.$

Decompositions on a small parameter $\varepsilon $ result in ansatz of type (%
\ref{finsbr}) with the coefficients subjected additionally to the conditions
(\ref{vaclcsoc}). The rest of nontrivial coefficients can be chosen to be
just those for a conformally transformed black ring metric (\ref{ctbrm}) on
a 6--d (pseudo) Riemannian space.

\subsection{Physical parameters for nonholonomic black rings}

\label{spbrp}A nonholonomic black ring configuration for a small $%
\varepsilon ,$ both for any zero and/or nonzero cosmological constant $%
\lambda ,$ can be characterized by physical parameters similar to those for
holonomic black ring configurations given by formulas (2.6)--(2.10) in
reference \cite{eev}. In the case of elliptic polarizations, we have to
change $\check{R}\rightarrow \check{r}$ (\ref{elconf}) and compute the
physical parameters for the metric (\ref{finsbr}) considering that a
corresponding observer is in a point in the v--part of spacetime and see a
conformally transformed and locally anisotropically polarized black ring
metric with deformed parameters on for direction $x^{2}$ (such an observer
is in nonholonmic/ ''N--adapted'' system of reference).

The corresponding effective locally anisotropic (ellipsoidal) mass,\newline
$M(\varepsilon ,x^{2}),$ angular momentum, $J(\varepsilon ,x^{2}),$
temperature, $T(\varepsilon ,x^{2}),$ angular velocity, $\varpi (\varepsilon
,x^{2}),$ and horizon area, $A_{H}(\varepsilon ,x^{2}),$ are given by
formulas%
\begin{eqnarray}
M &=&\frac{3\pi }{4C}\frac{\check{\lambda}}{1-\nu }\check{r}^{2}(\varepsilon
,x^{2}),\ \ J=\frac{\pi }{2C}\frac{\sqrt{\check{\lambda}(\check{\lambda}-\nu
)(1+\check{\lambda})}}{\left( 1-\nu \right) ^{2}}\check{r}^{3}(\varepsilon
,x^{2}),  \notag \\
T &=&\frac{1+\nu }{4\pi }\sqrt{\frac{1-\check{\lambda}}{\nu \check{\lambda}%
(1+\check{\lambda})}}\check{r}^{-1}(\varepsilon ,x^{2}),\ \varpi =\sqrt{%
\frac{\check{\lambda}-\nu }{\check{\lambda}(1+\check{\lambda})}}\check{r}%
^{-1}(\varepsilon ,x^{2}),  \notag \\
A_{H} &=&8\pi \frac{\nu ^{3/2}\sqrt{\check{\lambda}(1-\check{\lambda}^{2})}}{%
\left( 1-\nu \right) ^{2}(1+\nu )}\check{r}^{3}(\varepsilon ,x^{2}),
\label{phpar}
\end{eqnarray}%
where $\check{r}(\varepsilon ,x^{2})\rightarrow \check{R}$ for $\varepsilon
\rightarrow 0.$ Nevertheless, even in this limit such black ring
configurations do not transform always into the well known homogeneous black
ring solutions because the metric (\ref{finsbr}) may preserve its
nonholonomic character with certain nontrivial N--connection coefficients
(for instance, determined by certain nontrivial solitonic stationary
configurations).

We can say that a nonholonomically polarized black ring with physical
parameters (\ref{phpar}) mimics a generic ''off--diagonal'' metric
gravitational stationary configuration when an usual black ring is imbedded
self--consistently into a (pseudo) Finsler/Riemannian background with
gravitational polarizations like in an effective continuous media. For
simplicity, we have chosen the simplest ellipsoidal polarization with small
variable $\check{r}(\varepsilon ,x^{2})$ but it is possible to construct
more general classes of solutions when the constants $\nu $ and/ or $\check{%
\lambda}$ are also polarized. Such nonholonomic configurations with ring
topology can be also characterized by locally anisotropic physical
parameters (see some similar examples related to black torus and
torus--ellipsoid configurations in Ref. \cite{vsgg}).

\section{Concluding Remarks}

To summarize we have shown how black ring solutions can be constructed in
(pseudo) Finsler and nonholonomic (pseudo) Riemanian spacetimes. We used the
anholonomic deformation method of constructing exact solutions with generic
off--diagonal metrics in Einstein gravity and generalizations \cite%
{ijgmmp,vrflg,vncg,vsgg}. The Finsler--Einstein spacetimes with
anisotropically polarized constants and/or nonholonomic imbedding into
nontrivial backgrounds are interesting configurations to study from both
Finsler gravity and extra dimension gravity perspectives. Such objects can
be described by asymptotically flat gravitational configurations with a non
spherical horizon topology generated by nonholonomic deformations of
standard black ring metrics.

The first class of locally anisotropic stationary solutions was considered
for the Einstein equations with nontrivial constant in Finsler gravity and
six dimensional extension of general relativity. They were generated by
nonholonomic deformations of the black ring metric imbedded into
corresponding (pseudo) Finsler/ Riemannian manifolds. For the new classes of
Finsler--Einstein spacetimes, the dependence on the cosmological constant is
generic nonlinear and non--integrable. It is not clear what type of physical
interpretation such solutions may have in general even it is obvious that
they contain as a corresponding subclasses various types of black ring
solutions in braneworld gravity, string gravity with nontrivial cosmological
constant etc. Nevertheless, it is possible to extract physically important
solutions with nontrivial topology of horizon and gravitational
polarizations of cosmological constants using a procedure of extracting
solutions depending on a small parameter, for instance, characterizing small
''ellipsoidal'' deformations.

The second class of of Finsler type solutions are for vacuum configurations
defined by generic off--diagonal metrics and corresponding nonlinear and
linear connection structures. They also consist examples of
Finsler--Einstein spacetimes but, in general, they can not be generated in
the trivial limit of zero cosmological constant.

It might be possible to gain more insight on viability of gravity Finsler
type and extra dimension theories using properties of black hole/ring and
nonlinear wave solutions in such models and comparing them to similar ones
in Einstein gravity and string/brane gravity. Picturing such objects as
nonholonomic deformations of already studied physical gravitational systems
but with additional polarizations of physical constants, deformed symmetries
and imbedding, for instance, into solitonic backgrounds, we provide
realistic physical interpretations for a very general class of generic
off--diagonal metrics in general relativity and extra dimension gravity.

The question of stability of the Finsler like black ring configurations can
be solved explicitly for certain limits of small nonholonomic deformations
by using the analysis performed for usual \ holonomic black rings \cite{eev}
and black ellipsoid solutions \cite{vbep}. For general nonholonomic
transforms of a stable black ring metric, the target solution may be both
stable or unstable with an undetermined physical status.

As we discussed in the Introduction, the results of this work bridges the
gap between rather different directions of research: new methods of
constructing exact solutions in gravity theories, Finsler gravity and extra
dimension generalizations of general relativity. Our approach allows to
study solutions on tangent bundle and/or for nonholonomic manifold spacetime
models. The constructions on tangent bundles are related to theories with
violation of local Lorentz symmetry and so--called ''non--standard'' models
of gravity, see a conventional classification in \cite{vrflg}. An approach
to (pseudo) Finsler gravity and generalizations being compatible with the
modern paradigm of standard physics can be elaborated using the geometric
formalism of nonholonomic manifolds and associated nonlinear connection
structures.

It would be interesting to see whether the classes of solutions studied in
this work, and in the partner paper \cite{vfbh}, can be generalized for
noncommutative gravity \cite{vncg} and Ricci flow physical models \cite{vrf1}%
. Another very important extension of our results would be to encompass
gauge gravity and warped locally anisotropic geometries. Last, but not
least, it would be interesting to use the anholonomic deformation method to
construct cosmological solutions parametrized by generic off--diagonal
metrics and possessing generalized symmetries and anisotropic polarizations
of fundamental physical constants.

\vskip3pt

\textbf{Acknowledgement: } S. V. is grateful to R. Miron for very important
discussions and kind support.


\begin{thebibliography}{99}
\bibitem{perl} V. Perlick, Fermat principle in Finsler spacetimes, Gen. Rel.
Grav. \textbf{38} (2006) 365--380

\bibitem{mignemi} S. Mignemi, Doubly special relativity and Finsler
geometry, Phys. Rev. D \textbf{76} (2007) 047702

\bibitem{gibbons} G. W. Gibbons, J. Gomis and C. N. Pope, General very
special relativity is Finsler geometry, Phys. Rev. D \textbf{76} (2007)
081701

\bibitem{sindoni} L. Sindoni, The Higgs mechanism in Finsler spacetimes,
Phys.\ Rev. D \textbf{77} (2008) 1240009

\bibitem{skak} J. Skakala and M. Visser, Birefringence in pseudo--Finsler
spacetimes, J. Phys. Conf. Ser. \textbf{ 189} (2009) 012037 

\bibitem{vncg} S. Vacaru, Exact solutions with noncommutative symmetries in
Einstein and gauge gravity, J. Math. Phys. \textbf{46} (2005) 042503

\bibitem{vsgg} S. Vacaru, P. Stavrinos, E. Gaburov and D. Gon\c{t}a, \textit{%
Clifford and Riemann- Finsler Structures in Geometric Mechanics and Gravity,}%
\ Selected Works, Differential Geometry -- Dynamical Systems, Monograph 7
(Geometry Balkan Press, 2006);\newline
www.mathem.pub.ro/dgds/mono/va-t.pdf and gr-qc/0508023

\bibitem{vgont} S. Vacaru and D. Gontsa, Off---Diagonal Metrics and
Anisotropic Brane Inflation, hep-th/0109114; Contribution at the Conference
on Applied Differential Geometry -- General Relativity, Aristotle University
of Thessaloniki, School of Technology, Mathematics Devision, Thessaloniky,
Greece, June 27-July 1, 2001; Chapter 9 in Ref. \cite{vsgg}

\bibitem{lin} K. Lin and S.-Z. Yang, An inflationary solution of scalar
field in Finsler universe, Chin. Phys. Lett. \textbf{25} (2008) 2382--2384

\bibitem{chang} Z. Chang and X. Li, Modified Newton's gravity in Finsler
space as a possible alternative to dark matter hypothesis, Phys. Lett. B
\textbf{668} (2008) 453--456

\bibitem{kstav} A. P. Kouretsis, M. Stathakopoulos and P.\ C. Stavrinos, The
general very special relativity in Finsler cosmology, Phys. Rev. D\textbf{\
79 }(2009) 104011

\bibitem{ijgmmp} S. Vacaru, Parametric nonholonomic frame transforms and
exact solutions in gravity, Int. J. Geom. Methods. Mod. Phys. (IJGMMP)
\textbf{4} (2007) 1285-1334

\bibitem{vrflg} S. Vacaru, Finsler and Lagrange geometries in Einstein and
string gravity, Int. J. Geom. Methods. Mod. Phys. (IJGMMP) \textbf{5} (2008)
473-511

\bibitem{vfbh} S.\ Vacaru, Black Holes, Ellipsoids, and Nonlinear Waves in
Pseudo--Finsler Spaces and Einstein Gravity, arXiv: 0905.4401 [gr-qc]

\bibitem{ma1987} R. Miron and M. Anastasiei, Vector Bundles and Lagrange
Spaces with Applications to Relativity (Geometry Balkan Press, Bukharest,
1997); translation from Romanian of (Editura Academiei Romane, 1987)

\bibitem{ma} R. Miron and M. Anastasiei, The Geometry of Lagrange Spaces:\
Theory and Applications, FTPH no. \textbf{59} (Kluwer Academic Publishers,
Dordrecht, Boston, London, 1994)

\bibitem{mty} Y. Morisawa, S. Tomizawa and Y. Yasui, Boundary value problem
for black rings, Phys. Rev. D. \textbf{77} (2008) 064019

\bibitem{ps} A. A. Pomeransky and R.\ A. Sen'kov, Black ring with two
angular momenta, hep-th/0612005

\bibitem{er} R. Emparan and H. S. Real, Black Holes in Higher Dimensions,
Living. Rev. Rel. \textbf{11} (2008) 6

\bibitem{tz} P. K. Townsend and M. Zamklar, The first law of black brane
mechanics, Class. Quantum Grav. \textbf{18} (2001) 5269--5286

\bibitem{ar} D. Astefanesei and E. Radu, Quasilocal formalism and black ring
thermodynamics, Phys. Rev. D \textbf{73} (2006) 044014

\bibitem{ch} K. Copsey and G. Horowitz, The role of dipole charges in black
hole thermodynamics, Phys. Rev. D \textbf{73} (2006) 024015

\bibitem{rog1} M. Rogatko, Uniqueness theorem for stationary black ring
solution of sigma--models in five dimensions, Phys. Rev. D \textbf{77}
(2008) 124037

\bibitem{figueras} P. Figueras, A black ring with a rotating 2--sphere, JHEP
\textbf{0507} (2005) 039

\bibitem{eev} H. Elvan, R. Emparan and A. Virmani, Dynamics and stability of
black rings, JHEP \textbf{0612} (2006) 074

\bibitem{vbep} S. Vacaru, Perturbations and stability of black ellipsoids,
Int. J. Mod. Phys. D \textbf{12} (2003) 461--478

\bibitem{vs5dbh} S. Vacaru and D. Singleton, Warped solitonic deformations
and propagation of black holes in 5D vacuum gravity, Class. Quant. Gravity
\textbf{19 }(2002) 3583-3602

\bibitem{vrf1} S. Vacaru, Ricci Flows and Solitonic pp-Waves, Int. J. Mod.
Phys. A \textbf{21} (2006) 4899-4912

\bibitem{vsolh1} S. Vacaru, Curve Flows and Solitonic Hierarchies Generated
by Einstein Metrics,  Acta Applicandae Mathematicae [ACAP] \textbf{ 110 } (2010) 73--107 


\bibitem{vsolh2} S. Anco and S. Vacaru, Curve Flows in Lagrange-Finsler
Geometry, Bi-Hamiltonian Structures and Solitons, J. Geom. Phys. \textbf{59}
(2009) 79-103

\bibitem{bm} I. Bucataru and R. Miron, Finsler--Lagrange Geometry.
Applications to Dynamical Systems (Editure of Romanian Academy, 2007)
\end{thebibliography}
\end{document}